\documentclass{PoS}
\pdfoutput=1

\title{Excited mesons in a Bethe-Salpeter approach }

\ShortTitle{Excited mesons in a Bethe-Salpeter approach }

\author{\speaker{Andreas Krassnigg}\thanks{The author would like to acknowledge 
valuable discussions with G.~Eichmann, M.~Joergler, C.\,D.~Roberts, and M.~Schwinzerl.
This work was supported by the Austrian Science Fund \emph{FWF} under project no.\ P20496-N16 and
benefited from the computing resources of the Argonne National Laboratory LCRC.}\\
        Institut f\"ur Physik, Karl-Franzens-Universit\"at Graz, A-8010 Graz, Austria\\
        E-mail: \email{andreas.krassnigg@uni-graz.at}}

\abstract{%
In theoretical hadron physics mesons are a center of attention. Constructed in
a simpler way than baryons in the quark model, they still present a considerable
challenge if one aims at an understanding of all their aspects in terms of
quarks and gluons in the context of Quantum Chromodynamics, the quantum
field theory of the strong interaction. Complementary to (constituent-) quark models,
reductions of the Bethe-Salpeter equation, lattice QCD, and effective field theories,
the Dyson-Schwinger-equation approach has emerged as a well-suited formalism for the 
covariant study of hadron properties. In particular, radially excited mesons
exhibit a sensitivity to long-range strong-interaction physics. This sensitivity
has recently been studied with the help of the Bethe-Salpeter equation. Here
these studies are reviewed and continued together with an account of possible
future developments.%
}

\FullConference{8th Conference Quark Confinement and the Hadron Spectrum\\
		 September 1-6, 2008\\
		 Mainz. Germany}

\begin{document}

\section{Introduction}
Dyson-Schwinger equations (DSEs) provide a nonperturbative continuum approach to Quantum Chromodynamics (QCD)
(for recent reviews, see \cite{Fischer:2006ub,Roberts:2007jh}). A complete, simultaneous, and consistent solution 
of all DSEs results in the knowlegde of all Green functions of the theory. Mesons are studied within this approach
as quark-antiquark bound states in the Bethe-Salpeter equation (BSE) \cite{Smith:1969az}. In a
numerical study one usually resorts to a truncation of this infinite tower of coupled integral equations
by selecting only a few to be solved explicitly. One then incorporates assumptions about the solution of 
the rest of the equations via ans\"atze for those Green functions of the theory which are not solved
for explicitly. At this point Slavnov-Taylor- and Ward-Takahashi identities, representing the symmetries
of the underlying theory, serve as both restrictions and guides for such ans\"atze. In particular,
satisfaction of the axial-vector Ward-Takahashi identity (AVWTI) by appropriate construction of
the kernels of the quark-propagator DSE and the $q\bar{q}$ BSE correctly implements chiral symmetry
and its dynamical breaking in a truncation. A way to systematically do this in terms of a
nonperturbative scheme is known \cite{Munczek:1994zz,Bender:1996bb}. The lowest order in this scheme
is referred to as the ``rainbow-ladder'' (RL) truncation. While studies beyond RL have so far been mostly
of exploratory nature (see, e.\,g., \cite{Bhagwat:2004hn,Fischer:2005en,Matevosyan:2006bk,Fischer:2008wy} 
and references therein), various model interactions have been used to employ RL meson studies 
\cite{Munczek:1983dx,Jain:1993qh,Frank:1995uk,Burden:1997fq,Maris:1997tm,Alkofer:2002bp}.
The model of Ref.~\cite{Maris:1999nt} has been used in most detail to study
pseudoscalar- and vector-meson ground-state, in particular electromagnetic, properties 
(see \cite{Bhagwat:2006pu} and references therein). This interaction is also the basis for the present
calculation. While it was developed at and intended for light-quark masses, an extension to heavy quarks was
also explored \cite{Krassnigg:2004if,Maris:2005tt,Bhagwat:2006xi}, but ultimately only succeded recently 
\cite{Maris:2006ea}, since particular numerical care must be taken if one aims at reaching the $b$-quark 
mass (see also Sec.~\ref{details}).
Studies of radial meson excitations are another natural extension of the model of Ref.~\cite{Maris:1999nt}.
In particular, aspects of pseudoscalar and scalar meson excitations have been investigated
\cite{Holl:2004fr,Holl:2004un,Holl:2005vu,Krassnigg:2006ps,Bhagwat:2007rj}; here, vector-meson
radial excitations are added to the picture.
Note furthermore that the analogous approach to 
baryons as systems of three quarks is considerably more involved: there, intermediate steps have been
taken to allow for the same level of sophistication as in corresponding meson studies (for recent advances, see
\cite{Eichmann:2007nn,Eichmann:2008ef,Nicmorus:2008vb} and references therein).

\section{Details of the Approach and Numerical Technique}\label{details}
In RL truncation a meson with total $q\bar{q}$ momentum $P$ and relative $q\bar{q}$ 
momentum $k$ is described by the BS amplitude $\Gamma(k;P)$, the solution of the 
homogeneous, ladder-truncated $q\bar{q}$ BSE 
\begin{equation}\label{bse}
\Gamma(k;P) = -\frac{4}{3}\int^\Lambda_q \mathcal{G}((k-q)^2)\; D_{\mu\nu}^f(k-q) \;
\gamma_\mu \;S(q_+) \Gamma(q;P) S(q_-) \;\gamma_\nu \, \,,
\end{equation}
where Dirac and flavor indices have been omitted for simplicity and the factor 
$\frac{4}{3}$ comes from the color trace. $D_{\mu\nu}^f(p-q)$ is the free gluon 
propagator, $\gamma_\nu$ is the bare quark-gluon vertex, $\mathcal{G}((p-q)^2)$ 
is an effective running coupling, the (anti)quark momenta are $q_\pm = q\pm P/2$, 
and $\int^\Lambda_q=\int^\Lambda d^4q/(2\pi)^4$ represents a translationally invariant 
regularization of the integral, with the regularization scale $\Lambda$ \cite{Maris:1997tm}.  
$S(p)$ is the solution of the rainbow-truncated QCD gap equation 
\begin{equation}\label{dse}
S(p)^{-1}  =  (i\gamma\cdot p + m_q)+  \frac{4}{3}\int^\Lambda_q\! \mathcal{G}((p-q)^2) \; D_{\mu\nu}^f(p-q) 
\;\gamma_\mu \;S(q)\; \gamma_\nu \,, 
\end{equation}
where $m_q$ is the current-quark mass, and renormalization details have been omitted for simplicity.  
The quark propagator has the general form $S(p)^{-1}  =  i \gamma\cdot p \, A(p^2) + B(p^2)$.

In the $q\bar{q}$ BSE for equal-mass constituents in Euclidean momentum space a solution 
at $P^2=-M^2$ ($M$ is the bound-state mass) requires knowledge 
of the quark propagator at complex momenta whose squares 
lie inside a parabola defined by the complex points $(-M^2/4,0)$ and $(0,\pm M^2/2)$ (for
a more detailed discussion, see e.\,g.~App.~A of Ref.~\cite{Fischer:2008sp}). This 
can present a considerable numerical challenge at large quark 
masses (in practice for $m_q$ larger than the charm quark mass). In the 
present work a method based on the strategy outlined in \cite{Fischer:2005en} is used. 
The starting point is a switch of integration variable in Eq.~(\ref{dse}) from the quark momentum
used on the real axis to the gluon momentum, which is made possible here by the translationally
invariant regularization of all integrals (numerical checks on the real $p^2$ axis show
that errors from this procedure are indeed negligible compared to those already present from the integration
on the real axis). Integrating over the gluon momentum implies
that in order to solve Eq.~(\ref{dse}) for a complex value of $p^2$ one needs
to evaluate $\mathcal{G}$ for real arguments, while $S$ 
must be evaluated for complex arguments. As a consequence, simultaneous iteration
on a complex grid seems necessary. The same information as on a complex 2D grid can be obtained from 
a contour surrounding the region of interest, which suggests the use of the Cauchy formula. In Ref.~\cite{Fischer:2005en} 
the authors use a combination of contour integration and fitting: they close the parabolic contour at a
positive real value $q^2_{UV}$ large enough to ensure that values of $A$ and $B$ for arguments with larger
real parts than $q^2_{UV}$ are well described by fits to the corresponding UV behavior of the solutions on the
real axis. Then one needs to parameterize the parabolic contour and implement the necessary integrations
numerically. The Cauchy formula states that one can compute the value of a function $f(z)$ at a point $z_0$
inside a closed contour $\mathcal{C}$ from the values of the function on that contour using an $n$-point
quadrature as
\begin{equation}\label{cauchy}
f(z_0)=\frac{1}{2\pi i}\int_\mathcal{C}\frac{dz\;f(z)}{z-z_0}\approx
\frac{1}{2\pi i}\sum_{j=1}^n\frac{w_{jn}\;f(z_{jn})}{z_{jn}-z_0}\;,
\end{equation}
with $z_{jn}$ and $w_{jn}$ the $n$-point quadrature abscissae and weights on $\mathcal{C}$.
Such a straight-forward numerical implementation of the Cauchy formula is inaccurate as soon
as $z_0$ approaches the contour (for $z_0$ on the contour, the integral in Eq.~(\ref{cauchy}) is singular).
Hence, a different setup is in order.

Here, the implementation of this method in \cite{Fischer:2005en} is modified in two respects. 
First, the parabolic part of the contour is parameterized differently:
$z(t)=t^2/M_{max}^2-M_{max}^2/4+i\,t$, $dz(t)=(2t/M_{max}^2+i)dt$, $|t|<M_{max}\sqrt{q^2_{UV}+M_{max}^2/4}$.
Secondly, to avoid proximity of $z_0$ to the contour, the authors of Ref.~\cite{Fischer:2005en} modify their momentum
routing in the quark self energy and choose a routing such that both the arguments of the coupling and
the quark propagator in the self energy are complex. However, this is not necessary, since there is a
reliable numerical implementation of the Cauchy formula described in Ref.~\cite{Ioakimidis:1991io}.
There the authors use the Cauchy theorem to obtain a prescription for the evaluation of numerical 
correction terms for the integral in the Cauchy formula at every order in the computation of a function
and its derivatives. In particular, one has
\begin{equation}
f(z_0)\approx\frac{\left(\sum_{j=1}^n\frac{w_{jn}\;f(z_{jn})}{z_{jn}-z_0}\right)}{
\left(\sum_{j=1}^n\frac{w_{jn}}{z_{jn}-z_0}\right)}\;,
\qquad f'(z_0)\approx\frac{
\left(\sum_{j=1}^n\frac{w_{jn}\;f(z_{jn})}{(z_{jn}-z_0)^2}-f(z_0)\sum_{j=1}^n\frac{w_{jn}}{(z_{jn}-z_0)^2}\right)
}{\left(\sum_{j=1}^n\frac{w_{jn}}{z_{jn}-z_0}\right)}
\end{equation}
with similar formulae for higher derivatives.
As a result, the main source of numerical inaccuracy in this method is the
sampling of the function by the integration points chosen on the contour.

\section{Results and Discussion}
\begin{figure}
\includegraphics[width=0.62 \textwidth,clip=true]{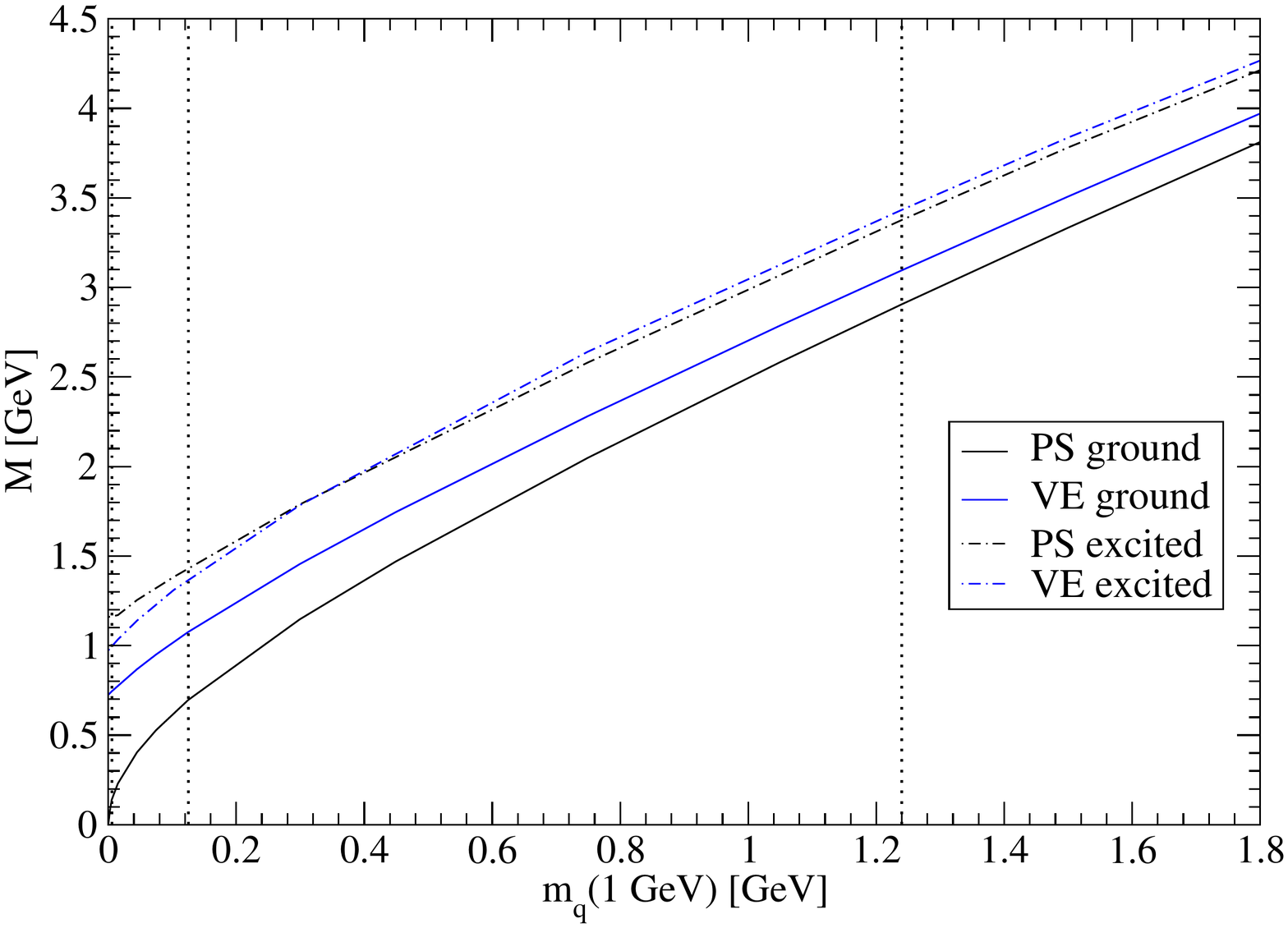}%
\includegraphics[width=0.62 \textwidth,clip=true]{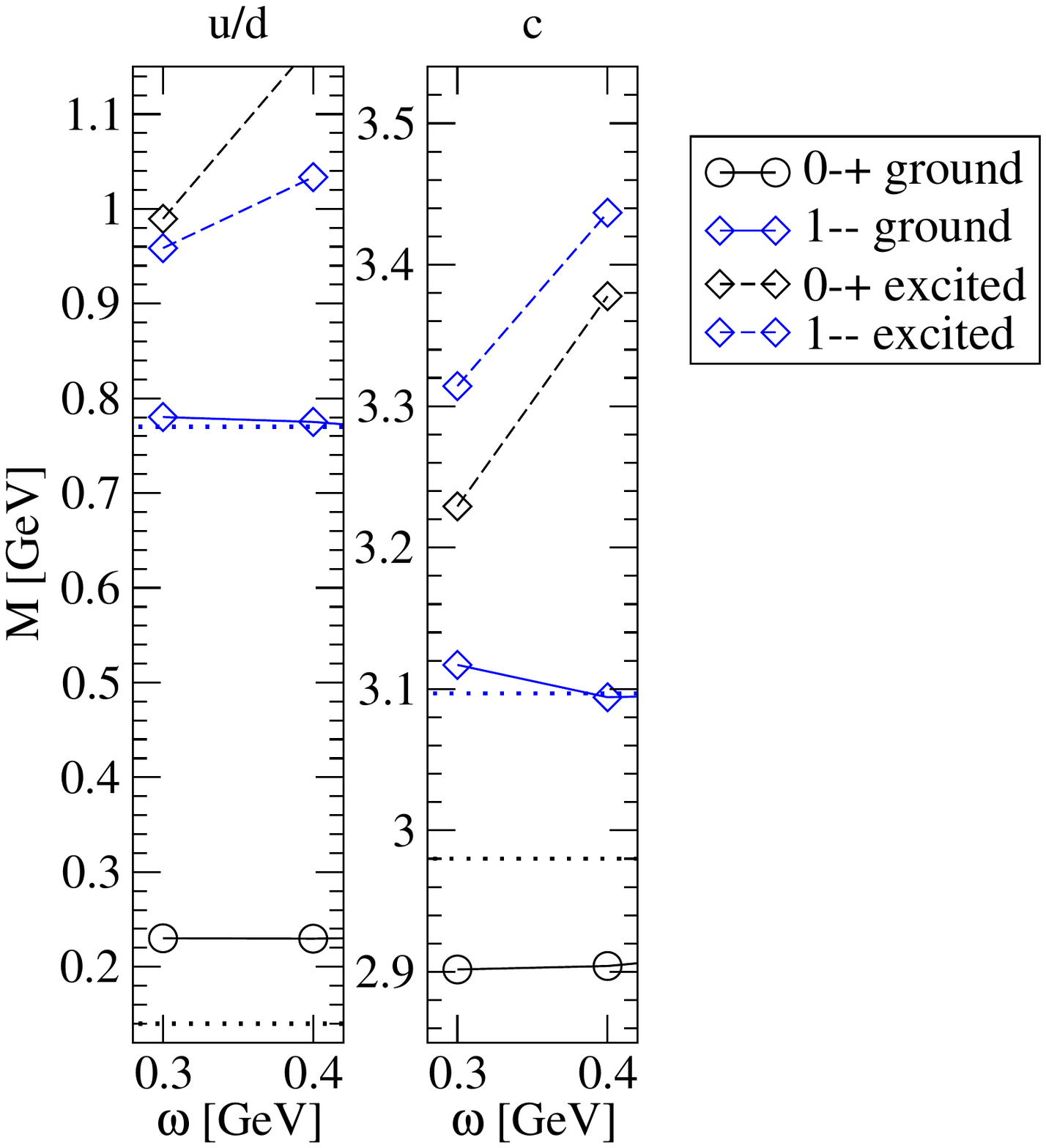}%
\caption{\textit{Left panel}: Pseudoscalar and vector meson masses as functions
of the current-quark mass, the vertical dotted lines denote the values of light, $s$ and $c$ quark masses; 
\textit{right panel}: Pseudoscalar and vector meson masses as functions
of the model parameter $\omega$ (see text), the horizontal dotted lines denote the corresponding
experimental values.\label{fig:1}}
\end{figure}

Using this setup, one can compute meson masses from the $q\bar{q}$ BSE. The left panel of Fig.~\ref{fig:1}
shows pseudoscalar- and vector-meson masses for the ground and first radially excited states, respectively,
as functions of the current-quark mass. The vertical dotted lines mark the positions of the light-, $s$-,
and $c$-quark masses, fixed by the condition that the vector state agrees with the experimental value.
Regarding radial excitations, an important observation is the following.
The model interaction of Ref.~\cite{Maris:1999nt} contains two terms; one makes sure that the coupling
has the correct UV behavior of the QCD running coupling and is not important for the following argument, 
while the other is an ansatz for the low- and intermediate-momentum ranges, i.\,e., it characterizes
the long-range part of the strong interaction. This term is of the form (for full detail, see \cite{Maris:1999nt})
$\mathcal{G}(k^2)/k^2\sim\frac{D}{\omega^4}\frac{k^2}{\omega^2} e^{-k^2/\omega^2}$, 
where $D$ and $\omega$ are free parameters. These were fitted to pion observables and the chiral condensate
in \cite{Maris:1999nt} for different
values of the parameter $\omega$, which can be interpreted as the inverse of an effective range of the strong
interaction in this model \cite{Holl:2005vu}. As a result, ground-state properties should not vary much as functions
%
of $\omega$, while properties of excited states should be more sensitive in this respect due to 
the reasonable assumption that they are larger than ground states. This is shown in the right panel
of Fig.~\ref{fig:1}, where the two cases of the light and charm quarks are exemplified; horizontal dotted 
lines mark the experimental values. It is evident that for radial excitations, even with the large dependence
on $\omega$ taken into consideration, the results considerably underestimate the experimental values (which lie 
well above the region plotted in the figure), in
particular in the vector case. While the overall qualitative picture is already coherent, further studies are
expected to yield quantitatively acceptable results after taking into account corrections to RL truncation.

In summary, a study of radial meson excitations in a Bethe-Salpeter approach immediately allows valuable insight
and qualitative studies, but further investigations are necessary before one can make reliable 
quantitative statements in this direction. Necessary improvements include adding corrections to 
the rainbow-ladder kernel as well as an explicit coupling to hadronic decay channels, i.\,e., a study
of meson resonances in the Bethe-Salpeter equation.

\bibliography{had_nucl_graz}

\end{document}